\documentclass[aps,prd,nofootinbib,preprint,amsmath,amssymb,floatfix]{revtex4-1}
\usepackage{color}
\usepackage{graphicx}
\usepackage{mathtools}
\usepackage{hyperref}
\usepackage{bm}% bold math
\usepackage{hyperref}
\hypersetup{colorlinks=true,% make the links colored
}

\def\lambdabar{\protect\@lambdabar}
\def\@lambdabar{%
\relax
\bgroup
\def\@tempa{\hbox{\raise.73\ht0
\hbox to0pt{\kern.25\wd0\vrule width.5\wd0
height.1pt depth.1pt\hss}\box0}}%
\mathchoice{\setbox0\hbox{$\displaystyle\lambda$}\@tempa}%
{\setbox0\hbox{$\textstyle\lambda$}\@tempa}%
{\setbox0\hbox{$\scriptstyle\lambda$}\@tempa}%
{\setbox0\hbox{$\scriptscriptstyle\lambda$}\@tempa}%
\egroup}

\begin{document}

\newcommand{\tc}{\textcolor}
\newcommand{\g}{blue}

\title{Rotation of the polarization plane in axion fields: application to neutron star polar cap regions}
% Enter your title between curly braces
\author{Iver H. Brevik}      % Enter your name between curly braces
\email{iver.h.brevik@ntnu.no}
\affiliation{Department of Energy and Process Engineering, Norwegian University of Science and Technology, N-7491 Trondheim, Norway}
\author{Moshe M. Chaichian}
\email{masud.chaichian@helsinki.fi}
\affiliation{Department of Physics, University of Helsinki, and Helsinki Institute of Physics,  P. O. Box 64, FI-00014 Helsinki, Finland}
\author{Tiberiu Harko}
\email{tiberiu.harko@aira.astro.ro}
\affiliation{Department of Physics, Babe\c s-Bolyai University, 1 Kog\u alniceanu Street, Cluj Napoca 400084, Romania,} 
\affiliation{Astronomical Observatory, 19 Cire\c silor Street, Cluj-Napoca 400487, Romania}
\author{Yuri N. Obukhov}
\email{obukhov@ibrae.ac.ru}
\affiliation{Theoretical Physics Laboratory, Nuclear Safety Institute,
Russian Academy of Sciences, B.Tulskaya 52, 115191 Moscow, Russia,}
\affiliation{L. D. Landau Institute for Theoretical Physics, Russian Academy of Sciences, 142432 Chernogolovka, Russia}

\date{\today}
% Enter your date or \today between curly braces

\begin{abstract}
We study observable manifestations of an axion field, focusing on possible polarizational effects for electromagnetic wave propagating in such a special magnetoelectric medium. The corresponding analysis is based on a geometric model of the Casimir type, in the framework of which the rotation angle of the polarization plane is derived to the lowest order. The results obtained are discussed for astrophysical conditions of a neutron star, where an existence of a locally inhomogeneous axion region is predicted in the polar cap. 
\end{abstract}
\maketitle
\newpage
%{
%  \hypersetup{linkcolor=blue}
%  \tableofcontents
%}

%%%%%%%%%%%%%%%%%%%%%%%%%%%%%%%%%%%%%%%%%%%%%%%%%%%%%%%%%%%%%%%%%%%%%%%%%%%%%%%%%%%%%%%%%%%%%%%%%%%%%%%%%
\section{Introduction}\label{secintro}
%%%%%%%%%%%%%%%%%%%%%%%%%%%%%%%%%%%%%%%%%%%%%%%%%%%%%%%%%%%%%%%%%%%%%%%%%%%%%%%%%%%%%%%%%%%%%%%%%%%%%%%%%

Several decades have gone since the axion hypothesis was introduced as an attempt to explain the $CP$ problem in high energy physics \cite{peccei77,peccei77d,sikivie83,weinberg78,preskill83,abbott83,dine83}. Among the very long list of papers in this area, a few of the more recent ones are listed in Refs.~\cite{peccei08,sikivie08A,sikivie14,sikivie21,chadhaday22,millar17,liu22,li91A,lawson19,kim19,sikivie03,mcdonald20,zyla20,carenza20,leroy20,brevik21a,oullet19,arza19,qiu17,dror21,brevik22a,fukushima19,tobar19,bae22,adshead20,patkos22,tobar22,derocco18,brevik22,brevik23,favitta23,brevik24,pandey24}. Up to this date, no experiment has so far been able to verify the existence of the axion particle. Popular experimental methods are the haloscope type of experiments pioneered by Sikivie \cite{sikivie83,sikivie08A,sikivie14,sikivie21} (an extensive treatment of the dielectric haloscope can be found in Ref.~\cite{millar17}), and the axion dark-matter birefringent cavity experiment \cite{pandey24}. Somewhat surprisingly, the range of mass of the axion (or axion-like) is at present exceedingly wide, from about $10^{-20}\,{\rm eV}/c^2$ to $10^{-2}\,{\rm eV}/c^2$ \cite{chadhaday22,pandey24}.

The current interest in axion physics is however not restricted to observations only. From a formal point of view, the  axion electrodynamics is characterized by a constitutive tensor with non-diagonal terms responsible for the magnetoelectric effects \cite{Post,ODell,Birk}. That means in general nonreciprocal topological matter, with a variety of applications (cf., for instance, the review \cite{asadchy20}). Another extensive treatise on electromagnetic non-reciprocity is Ref.~\cite{lu21}, where, among other things, an analysis of the Casimir effect for such materials is explored. If the constitutive tensor contains imaginary non-diagonal terms, one has to do  with electromagnetic chiral materials \cite{mun20}, implying a rotation of the polarization plane for electromagnetic waves \cite{jiang19}.

Recent investigations of Noordhuis {\it et al.} \cite{noordhuis23,noordhuis24} and others have demonstrated the possible occurrence of strong local inhomogeneous axion regions in the polar cap of a neutron star. These regions, referred to as {\it vacuum gaps}, are characterized by static magnetic fields $B_0 \sim 10^8\,$T (=$10^{12}\,$G) directed normally outwards from the polar surface (magnetic dipole), together with static electric fields $E_0 \sim 10^{-6}cB_0~$ in the same direction (electric dipole). An enormous increase  of rate of axion production, up to order $10^{50}$,  is predicted in these small vacuum gap regions in the polar cap. These features make it of interest to reconsider phenomena such as polarization plane rotation under both weak and strong axion field  conditions. This is the main theme of the present work.

We begin by surveying the very peculiar antenna property of dielectric materials (especially conductors), which shows the need of having very strong magnetic fields in order to achieve measurability. After a concise introduction to the axion electrodynamics in Sec.~\ref{fieldeqs1} and \ref{fieldeqs2}, we present in Sec.~\ref{model} the general form for electromagnetic waves in the axion environment, in standard form as well as in a physically instructive hybrid form showing non-reciprocity of the axion fluid, and we calculate the polarization rotation. As expected, the rotation is well defined in the case of weak axion fields, even if they are  stronger than those  in the mean Universe. For very strong fields, this kind of perturbative theory breaks down, however.  A noteworthy general property of the rotation of the polarization plane is that it can only occur when the axion cloud is varying in space or time. We limit ourselves to  only space variations in this work.

Finally, as  an application we discuss  the physical picture of the local `gap' regions proposed by Noordhuis and others in the polar regions of a neutron star. The  reason for the occurrence of these gaps is plasma effects in the cloud. To get an idea of the time scales involved, we  calculate the filling time for surrounding axions flowing into an initial gap. It turns out that the typical filling time is a moderate number of nanoseconds, what appears to be a reasonable result.

%%%%%%%%%%%%%%%%%%%%%%%%%%%%%%%%%%%%%%%%%%%%%%%%%%%%%%%%%%%%%%%%%%%%%%%%%%%%%%%%%%%%%%%%%%%%%%%%%%%%%%%%%
\subsection{Search for axion}
%%%%%%%%%%%%%%%%%%%%%%%%%%%%%%%%%%%%%%%%%%%%%%%%%%%%%%%%%%%%%%%%%%%%%%%%%%%%%%%%%%%%%%%%%%%%%%%%%%%%%%%%%

Several experiments and proposals for experiments have recently been carried out or proposed for how to measure the pseudoscalar axion field, whose amplitude is conventionally called $a(x)$ with $x$ meaning $(t, {\bm x})$. The most well known example is the haloscope idea, suggested by Sikivie as mentioned above. These experiments have turned out to be exceedingly challenging. Let us here give a short survey of another interesting idea, namely to exploit the `antenna' property of a dielectric surface separating a half-space with high refractive index from a half-space with low refractive index (in practice a vacuum) \cite{millar17,brevik23}. Let the surface be the plane $x=0$, and let the right medium 2 be a perfect metal with infinite refractive index $n_2 \rightarrow \infty$ while the left medium 1 is a vacuum with $n_1=1$. The electromagnetic wave propagates in the $x$ direction. A strong static magnetic field $B_0$ is assumed  in the  $z$ direction. The antenna action at the surface reflects that the boundary conditions become changed because of the axions. We assume first  that the  axion field is uniform in space and  varies harmonically in time,
\begin{equation}
a(t) = a_0 \cos \omega_a t,
\end{equation}
corresponding to the energy density
\begin{equation}
\rho_a(t) = \frac{1}{2} {\dot a}^2(t) +\frac{1}{2}\omega_a^2 a^2(t),
\end{equation}
with $\omega_a$ the eigenfrequency. As axions are believed to vary relatively slowly (velocities $\leq c/3$), we may put the frequency equal to the axion mass $\omega_a = m_ac^2/\hbar$, which we will assume for definiteness to be equal to
\begin{equation}
m_a = 10^{-5}~ \rm{eV}/c^2.
\end{equation}
As the mean energy density  $\langle \rho_a \rangle = \frac{1}{2}m_a^2a_0^2$ is commonly assumed to be about 0.045 J/m$^3$, we  find that the axion amplitude is about
\begin{equation}
a_0 = 260~\rm{eV}.
\end{equation}
The boundary condition across the surface $x=0 $ mentioned above gives the following expression for the amplitude of the photon field, called $ E_{1\gamma}$, on
 the left side,
\begin{equation}
E_{1\gamma} =  -\,{\frac{E_0}{n_1}}\left({\frac{1}{n_2}} - {\frac{1}{n_1}}\right), \label{antenne}
\end{equation}
where $E_0$ means
\begin{equation}
E_0= \theta_0\,cB_0. \label{4}
\end{equation}
Here $\theta_0$ is the value of the axion-photon coupling,
\begin{equation}
\theta_0 = g_{a\gamma\gamma}a_0.
\end{equation}
We will assume the conventional value $g_{a\gamma\gamma}= 10^{-12}~$GeV$^{-1}$ for the coupling constant. Thus
\begin{equation}
\theta_0 = 2.6\times 10^{-19}.
\end{equation}
In SI units, we observe that the ratio of the energy densities $W_1$ and $W_2$  on the two sides is related as the square of the ratio of the axion field values,
\begin{equation}
W_1 = \theta_0^2\, W_2.
\end{equation}
On the right side 2  we obtain, assuming $B_0 = 10~$T, $W_2 = {\frac 1{2\mu_0}}B_0^2 = 4.0\times 10^7~$J/m$^3$. On the left side 1 we associate $W_1$ with the  electromagnetic wave energy density, $W_1 = {\frac {\varepsilon_0}2}E_{\rm rms}^2$, where rms means the root-mean-square of the field. Using that $E_{\rm rms}=E_0/\sqrt{2}$, we then see that  $W_1$ is reduced significantly, $W_1 = 2.7\times 10^{-30}~\rm{J/m}^3$. Multiplying  with a factor $c$ to get the energy flux intensity $S$ for the electromagnetic wave emitted to the left, we obtain
\begin{equation}
S = 8\times 10^{-22}~\rm{W/m}^{2}.
\end{equation}
This is very small; it is  comparable more or less with the energy flux density from a deliverance of a 1 eV photon per day (what corresponds to about $10^{-24}~$W/m$^2$).

In spite of these obvious experimental difficulties, there have been experiments done by the BREAD Collaboration attempting to verify the existence of this kind of photons (`dark photon dark matter' tests \cite{liu22,knirck24}). An ingenious parabolic  reflector was constructed, so as to focus the emitted photons onto a horn antenna.  Although the experimental data reported in Ref.~\cite{knirck24} did not show evidence for the desired kind of photons, the results were useful in providing strengthened constraints on the dark photon-photon mixing parameters.

These circumstances make it natural to leave terrestrial conditions and look  into the outer space instead. Recent investigations have indicated that there exists a mechanism for productions of axions  in the polar region of pulsar magnetospheres \cite{prabhu21,noordhuis23,noordhuis24}. The proposed mechanism is that relativistic  axions can be produced  in local vacuum regions near the polar caps of neutron stars from the spacetime oscillations of $\bm{E}\cdot \bm{B}$. The neutron star effective frequency causing the oscillations  is typically set to be $\Omega/2\pi \approx 1~$Hz.  This gives quite an important new insight into the physical conditions of the polar regions. In the mentioned papers it is derived that with the mentioned strong magnetic field of $B_0 = 10^8\,$T (=$10^{12}\,$G), and also a strong radially directed electric field $E_0 =6 \times 10^{-6}cB_0$, the production rate  of axions in the polar cap may be enormous, of order $N \sim 10^{50}$ per second.  This is quite a dramatic new observation, as all active neutron stars are expected to be surrounded by dense action clouds. One expects the local densities of axions in these clouds to be of order $10^{28}\,$GeV/m$^{3}$, thus enormously larger than the average value of $3.5\times 10^{5}$\,GeV/m$^{3}$ commonly assumed in the outer space. Our goal in the following will be to make use of this observation to calculate the rotation of the electromagnetic polarization plane when an axionic fluid propagates in a parallel-plate Casimir cavity. We will assume the same strong static magnetic field as reported in the mentioned papers, $B_0 = 10^{8}\,$T, somewhat below the critical Schwinger limit $B_c = 4.41\times 10^{9}\,$T. The present investigation is a follow-up of the calculation that we made before, under more moderate conditions. For completeness we will take the axion fluid to be nonreciprocal, meaning that the constitution tensor (in the hybrid formulation) contains non-diagonal real terms. Although that assumption is so far theoretical only, it is nevertheless of interest in view of the large attention that is at present devoted to that kind of materials.   In general, two  different reasons for the rotation of the polarization plane are:

\noindent 1) Rotation because of spatial inhomogeneity of the axion field;

\noindent 2) rotation because of time variation of the same field.

We will not consider the second option in the present paper. It is notable that the mentioned non-reciprocity of the axion-electromagnetic field in the hybrid formulation does not cause any rotation of the polarization plane.

We now turn to the field equations, in two different versions.

%%%%%%%%%%%%%%%%%%%%%%%%%%%%%%%%%%%%%%%%%%%%%%%%%%%%%%%%%%%%%%%%%%%%%%%%%%%%%%%%%%%%%%%%%%%%%%%%%%%%%%%%%
\section{Field equations: conventional form}\label{fieldeqs1}
%%%%%%%%%%%%%%%%%%%%%%%%%%%%%%%%%%%%%%%%%%%%%%%%%%%%%%%%%%%%%%%%%%%%%%%%%%%%%%%%%%%%%%%%%%%%%%%%%%%%%%%%%

In the presence of external electric and magnetic fields,  the constitutive relations in the rest inertial  system are
\begin{equation}
{\bm D}=\varepsilon\varepsilon_0 {\bm E}, \qquad {\bm B}= \mu\mu_0 {\bm H},
\end{equation}
with $\varepsilon$ and $\mu$ assumed constants. In SI unit system, the extended Maxwell equations of the axion electrodynamics become
\begin{equation}
{\bm \nabla} \cdot \bm{D}= \rho - \sqrt{\frac {\varepsilon_0}{\mu_0}}\bm{B}\cdot\bm{\nabla}\theta, \label{5}
\end{equation}
\begin{equation}
{\bm \nabla} \times \bm{H}= {\bm J} + \frac{\partial {\bm D}}{\partial t} + \sqrt{\frac {\varepsilon_0}{\mu_0}}
\left(\frac{\partial \theta}{\partial t}\bm{B} + \bm{\nabla}\theta\times\bm{E}\right), \label{7}
\end{equation}
\begin{equation}
{\bm \nabla} \cdot \bm{B}=0, \label{8}
\end{equation}
\begin{equation}
{\bm \nabla} \times \bm{E} = -\,\frac{\partial {\bm B}}{\partial t}. \label{9}
\end{equation}
As these equations are derived from a relativistic invariant Lagrangian, they will hold in any inertial system just as the ordinary Maxwell equations do (although it should be observed  that $\varepsilon$ and $\mu$ have their simple physical meaning in the rest system only). A covariant version of this phenomenological electrodynamics is achieved by introducing {\it two} field tensors, the usual form $F_{\mu\nu}= \partial_\mu A_\nu - \partial_\nu A_\mu$, plus the response tensor $H_{\mu\nu}$ \cite{moller72} (see also \cite{Birk}).

The  governing equations for the fields are
\begin{eqnarray}
\nabla^2 {\bm E} - {\frac {\varepsilon\mu}{c^2}} \frac{\partial^2 {\bm E}}{\partial t^2} &=&
\bm{\nabla}(\bm{\nabla} \cdot \bm{E}) +  \mu\mu_0\frac{\partial {\bm J}}{\partial t}
+ {\frac {\mu}{c}}\frac{\partial}{\partial t}\left[\frac{\partial \theta }{\partial t}{\bm B}
+ \bm{\nabla}\theta\times \bm{E}\right], \label{10}\\
\nabla^2 {\bm B} - {\frac {\varepsilon\mu}{c^2}}\frac{\partial^2 {\bm B}}{\partial t^2} &=&
- \mu\mu_0\bm{\nabla}\times\bm{J} - {\frac {\mu}{c}}\bm{\nabla}\times\left[\frac{\partial \theta}{\partial t}{\bm B}
+ \bm{\nabla}\theta\times \bm{E}\right]. \label{11}
\end{eqnarray}
The field equations above are complicated in the sense that they contain the second order derivatives of $\theta$. These can  be removed if we consider the approximation where $\partial^2 \theta /\partial t^2$ and $\partial \bm{\nabla}\theta /\partial t$ are small or zero.  We shall assume this limitation  in the following. Then, the field equations take the reduced forms
 \begin{eqnarray}
\nabla^2 {\bm E}- {\frac {\varepsilon \mu}{c^2}}\frac{\partial^2 {\bm E}}{\partial t^2 } &=&
\bm{\nabla}(\bm{\nabla} \cdot \bm{E}) + \mu\mu_0 \frac{ \partial {\bm J}}{\partial t}
+ {\frac {\mu}{c}}\left[ \frac{\partial \theta}{\partial t}\frac{\partial {\bm B}}{\partial t} +
\bm{\nabla}\theta\times\frac{\partial {\bm E}}{\partial t} \right], \label{12}\\
\nabla^2{\bm B} - {\frac {\varepsilon\mu}{c^2}} \frac{\partial^2{\bm B}}{\partial t^2} &=& - \,\mu\mu_0
\bm{\nabla} \times\bm{J} - {\frac {\mu}{c}}\left[\frac{\partial\theta}{\partial t}\bm{\nabla} \times \bm{B}
+ (\bm{\nabla}\theta)\bm{\nabla}\cdot\bm{E} -(\bm{\nabla}\theta\cdot\bm{\nabla}){\bm E}\right]. \label{13}
 \end{eqnarray}
The following equation is also useful:
\begin{equation}
\bm{\nabla}(\bm{\nabla} \cdot {\bm E}) = \frac{1}{\varepsilon}\bm{\nabla}\left[
\rho/\varepsilon_0 - c\bm{B}\cdot\bm{\nabla}\theta\right]. \label{extra}
\end{equation}

%%%%%%%%%%%%%%%%%%%%%%%%%%%%%%%%%%%%%%%%%%%%%%%%%%%%%%%%%%%%%%%%%%%%%%%%%%%%%%%%%%%%%%%%%%%%%%%%%%%%%%%%%
\section{Alternative hybrid form}\label{fieldeqs2}
%%%%%%%%%%%%%%%%%%%%%%%%%%%%%%%%%%%%%%%%%%%%%%%%%%%%%%%%%%%%%%%%%%%%%%%%%%%%%%%%%%%%%%%%%%%%%%%%%%%%%%%%%

This alternative form for Maxwell's equations  is worth attention, as it shows the axionic dependent terms explicitly, as source terms. With that, one recognizes the formal similarity with  ordinary electrodynamics of the  electro-optic group of media. A further advantage of this reformulation of the governing equations is that the  boundary conditions at a dielectric surface follow in a very transparent way.

We introduce two new field tensors, here called ${\bm D}'$ and ${\bm H}'$,
\begin{equation}
{\bm D}'= \varepsilon\varepsilon_0 {\bm E}+\sqrt{\frac {\varepsilon_0}{\mu_0}}\theta{\bm B},
\qquad {\bm H}'= {\bm H} - \sqrt{\frac {\varepsilon_0}{\mu_0}}\theta {\bm E}.
\end{equation}
When written as
\begin{equation}
\left(\begin{array}{ll}
{\bm D}' \\
{\bm H}'
\end{array}\right)
= \sqrt{\frac {\varepsilon_0}{\mu_0}}
\left( \begin{array}{cc}
\varepsilon/c & \theta \\
-\theta & c/\mu
\end{array}
\right)
\left(\begin{array}{ll}
{\bm E} \\
{\bm B}
\end{array}
\right), \label{20}
\end{equation}
it is seen how ${\bm D}', {\bm H}'$  relate to the response tensor $H^{\mu\nu}$ in the matter as we mentioned above, and not to the original field tensor $F_{\mu\nu}$. In terms of the new fields, the Maxwell equations get formally the same appearance  as in usual electrodynamics,
\begin{equation}
\bm{\nabla}\times\bm{H}' = {\bm J}+ \frac{\partial {\bm D}'}{\partial t}, \qquad {\bm\nabla}\cdot\bm{D}' = \rho,
\end{equation}
\begin{equation}
\bm{\nabla}\times\bm{E} = -\,\frac{\partial {\bm B}}{\partial t},\qquad \bm{\nabla}\cdot\bm{B} =0.
\end{equation}
Notice that the fields $\bm E$ and $\bm B$ are the same as in our first formulation above.

Since $\theta$ is assumed real in axionic theory, the non-diagonal terms in  the constitution matrix (\ref{20}) are real. This property refers to configuration space. In Fourier space, the non-diagonal terms would be imaginary \cite{mun20}. This is because the non-diagonal terms refer to quantities having opposite parities (${\bm D}'$ and ${\bm H}'$, respective $\bm E$ and  $\bm B$). The real and the imaginary parts of the non-diagonal terms are connected via the Kramers-Kronig formulas.

The  boundary conditions at a dielectric boundary in the hybrid formulation were considered in detail in Ref.~\cite{brevik23}.

It is worthwhile to mention that as a manifestation of the magnetoelectric effect in condensed matter physics, the axion was actually measured for multiferroics \cite{PLA:2005}. The polarization rotation of light in magnetoelectric media was also considered in \cite{PLA:2005}.

%%%%%%%%%%%%%%%%%%%%%%%%%%%%%%%%%%%%%%%%%%%%%%%%%%%%%%%%%%%%%%%%%%%%%%%%%%%%%%%%%%%%%%%%%%%%%%%%%%%%%%%%%
\section{Model}\label{model}
%%%%%%%%%%%%%%%%%%%%%%%%%%%%%%%%%%%%%%%%%%%%%%%%%%%%%%%%%%%%%%%%%%%%%%%%%%%%%%%%%%%%%%%%%%%%%%%%%%%%%%%%%

%%%%%%%%%%%%%%%%%%%%%%%%%%%%%%%%%%%%%%%%%%%%%%%%%%%%%%%%%%%%%%%%%%%%%%%%%%%%%%%%%%%%%%%%%%%%%%%%%%%%%%%%%
\subsection{Casimir-type setup, and the rotation of the polarization plane}
%%%%%%%%%%%%%%%%%%%%%%%%%%%%%%%%%%%%%%%%%%%%%%%%%%%%%%%%%%%%%%%%%%%%%%%%%%%%%%%%%%%%%%%%%%%%%%%%%%%%%%%%%

In the following, by considering the potential applications of our results for the case of neutron star physics, we investigate the properties of an axion-dominated area that may form in a polar cap of neutron stars \cite{prabhu21,noordhuis23,noordhuis24}. We further assume that this area is thin, as compared to the extension of the compact, and it contains a negligible mass. If the thickness $dr$ of the axion cloud influenced zone is much smaller than the radius of the star, $dr\ll R$,  one can use a Cartesian system of reference  even for the description of the gravitational/physical properties of a spherical object. A standard example in this context is the equation of the hydrostatic equilibrium as applied to a thin atmosphere, which can be written as $dP/dz=\rho (z) g(z)$, where $P$ is the pressure, $\rho (z)$ is the density, $g(z)$ is the surface gravity, and $z$ is the depth coordinate. Hence in the approximation of the thin axion dominated region we model the axion dominated region  as two plan-parallel slabs (plates) atop the neutron star, representing a standard approximation in the study of the stellar structure \cite{crust}.  Moreover, following \cite{noordhuis23,noordhuis24} we will assume that the boundaries of the plates can be modeled as having conducting properties. It should also be noted that the electric conductivity of a neutron star surface is extremely high, and it is determined by the flow  of a degenerate relativistic electron gas through the crystalline lattice of atomic nuclei. The high conductivity of the neutron star surface, determined  by electron-ion scattering (phonons),  is generally anisotropic due to the presence of magnetic fields, and strongly depends on the impurities in the surface \cite{cond}.

Hence, we will consider the in the following the two-plate model, usually also employed in Casimir physics, namely  two conducting parallel  large plates situated at $z=0$ and $z=L$. In the cavity there is an axion-influenced electromagnetic field. As in  Ref.~\cite{brevik24} we  will assume that the axion field is $\theta = 0$ at $z=0$ and is linearly increasing with increasing $z$. It means that the vertical gradient of $\theta$ is constant,
\begin{equation}
\bm{\nabla}{\theta} = \beta \hat{\bm{e}}_z.
\end{equation}
We  assume there is a strong, static, magnetic field ${\bm B}_0$  between the plates,
\begin{equation}
{\bm B}_0 = B_0\hat{\bm{e}}_z,
\end{equation}
and we  also include a strong, static, electric field ${\bm E}_0$. For simplicity we will take it to be collinear with ${\bm B}_0$,
\begin{equation}
{\bm E}_0 = E_0\hat{\bm{e}}_z.
\end{equation}

This assumption is supported by the   model of Noordhuis et al. \cite{noordhuis23,noordhuis24}: they assume  that both the magnetic and the electric field lines emerge in an orthogonal direction from the cap region of the neutron star.

Since the width $L$ is finite, the system is not translationally  invariant in the $z$ direction, but it is invariant in the  $\bm{x}_\bot = (x, y)$ directions.

We assume the following ansatz for the fields (with $\bm{k}_\bot = (k_x, k_y)$ and ${\bm k}_\perp^2 = k_x^2+k_y^2$):
\begin{equation}\label{ansatzE}
\bm{E}(t,\bm{x}) = \bm{E}(z)e^{i\Phi}, \qquad \Phi= \bm{k}_\bot \cdot \bm{x}_\perp - \omega t,
\end{equation}
 and consider the reduced governing equation (\ref{12}) for $\bm E$. As we limit ourselves to the electromagnetic subsystem (and not the governing equation for the axions), we can put $\rho = {\bm J}=0.$ We start from the $z$ equation,
 \begin{equation}
E_z''(z)+ \lambda^2 E_z(z) = 0,
\end{equation}
where $\lambda^2$ is defined as
\begin{equation}
\lambda^2 = {\frac {\varepsilon\mu\omega^2}{c^2}} - {\bm k}_\perp^2.
\end{equation}
The general solutions are thus of the form $\sin \lambda z$ and $\cos \lambda z$. If both these should be included, depends on the nature of the boundaries $z=0$ and $z=L$.

Recalling the standard accepted model, the outer region of a neutron star is a magnetosphere which is densely filled with the electron-positron plasma, the presence of which is important for the emission of coherent radio waves and for the stability of the magnetosphere \cite{prabhu21}. The vacuum gaps are small regions in the polar cap between the neutron star's surface and the plasma-filled zone of the magnetosphere, see, e.g., Fig. 1 in  \cite{noordhuis23}. Following Noordhuis et al. \cite{noordhuis23,noordhuis24} both boundaries -- star's surface and the plasma-filled zone -- are modeled as perfect conductors.

Accordingly, we assume here conducting boundaries, so that the only mode solution is the TM mode
\begin{equation}\label{Ez}
E_z(z)= {\mathcal C}_n \cos \lambda_nz, \qquad \lambda_n= {\frac {n\pi}L}, \qquad n=1,2,3,\dots,
\end{equation}
with ${\mathcal C}_n$ a constant.

We introduce a dimensionless  parameter $\xi_n$,
\begin{equation}
\xi_n = \frac{\beta \omega}{\lambda_n^2c}.
\end{equation}
If $\xi_n \ll 1$, the influence from the axions is weak. For general values of $\xi_n$, the simplified  equations (\ref{12}) and (\ref{13}) in the $x$ and $y$ directions can be written as
\begin{align}
E_x''(z) + \lambda_n^2 E_x(z) &= i\mu \lambda_n^2 \xi_n E_y(z), \label{exligning}\\
E_y''(z) + \lambda_n^2 E_y(z) &= -i\mu \lambda_n^2 \xi_n E_x(z). \label{exligning1}
\end{align}
Notice that the $\bm{\nabla}(\bm{\nabla}\cdot\bm{E})$ does not contribute, in view of Eq.~(\ref{extra}).

The two differential equations above can be rewritten as
\begin{align}
{\cal{E}}_{(+)}'' + \lambda_n^2(1-\mu\xi_n){\cal{E}}_{(+)} &= 0, \\
{\cal{E}}_{(-)}'' + \lambda_n^2(1+\mu\xi_n){\cal{E}}_{(-)} &= 0,
\end{align}
where ${\cal{E}}_{(+)}= E_x+iE_y$ and ${\cal{E}}_{(-)}=E_x-iE_y$. The solutions are thus
\begin{align}
E_x &= {\cal{E}}_{(-)}^0 \exp{(i\omega_{(-)}z)} + {\cal{E}}_{(+)}^0 \exp(i\omega_{(+)}z), \label{37}\\
E_y &= i{\cal{E}}_{(-)}^0 \exp{(i\omega_{(-)}z)} -i{\cal{E}}_{(+)}^0 \exp(i\omega_{(+)}z), \label{38}
\end{align}
where ${\cal{E}}_{(+)}^0$ and ${\cal{E}}_{(-)}^0$ are constants, and
\begin{align}
\omega_{(+)}^2 &= \lambda_n^2(1-\mu\xi_n),\\
\omega_{(-)}^2 &= \lambda_n^2(1+\mu\xi_n).
\end{align}
In order to streamline the discussion, let us focus on the lowest mode $n=1$ only; so that $\lambda_n \rightarrow \lambda_1 \equiv \lambda$ and $\xi_n \rightarrow \xi_1 \equiv \xi$. Then assuming weak coupling, $\xi \ll 1$, for the simplest initial conditions, eqs. (\ref{Ez}), (\ref{37}) and (\ref{38}) yield the final solution for the vector $\bm{E}(z)$ in (\ref{ansatzE}):
\begin{equation}
\bm{E}(z) = \cos(\lambda z)\left\{{\mathcal A}\cos\phi(z)\hat{\bm{e}}_x
+ {\mathcal A}\sin\phi(z)\hat{\bm{e}}_y + {\mathcal C}\,\hat{\bm{e}}_z \right\}.\label{Esol}
\end{equation}
Here ${\mathcal C}$ and ${\mathcal A}$ are integration constants, and the angle
\begin{equation}\label{maxrotation}
\phi(z)  = -\,{\frac{\mu \lambda \xi}{2}}\,z,
\end{equation}
measures the rotation of the polarization plane (recall that $\xi = \beta \omega/c\lambda^2, \, \beta = d\theta/dz$).

In principle, this rotation is a measurable quantity.

%%%%%%%%%%%%%%%%%%%%%%%%%%%%%%%%%%%%%%%%%%%%%%%%%%%%%%%%%%%%%%%%%%%%%%%%%%%%%%%%%%%%%%%%%%%%%%%%%%%%%%%%%
\subsection{Numerical estimates}
%%%%%%%%%%%%%%%%%%%%%%%%%%%%%%%%%%%%%%%%%%%%%%%%%%%%%%%%%%%%%%%%%%%%%%%%%%%%%%%%%%%%%%%%%%%%%%%%%%%%%%%%%

Let us first consider the mean Universe, recalling from above our chosen values
\begin{equation}
\langle \rho_a \rangle = \frac{1}{2}m_a^2a_0^2 = 0.045\, {\frac {\rm J}{{\rm m}^3}},% 0.45~ {\rm{ erg/cm}^3},
\qquad m_a= {\frac {\hbar\omega_a}{c^2}} = 10^{-5}~{\frac {\rm{eV}}{c^2}}, \qquad a_0 = 260~\rm{eV}.
\end{equation}
Recall that $a_0$ is the axion field value. It corresponds to the dimensionless
coupling constant  parameter $\theta_0= g_{a\gamma\gamma}a_0= 2.6\times 10^{-19}$.   

It is of interest to check the magnitude of the parameter $\xi = \beta \omega_a/c\lambda$, and also to give an estimate of  $\phi(z)$. As $\beta= g_{a\gamma\gamma}a_0/L = \theta_0/L$ and $\lambda = \pi/L$,  the magnitude of the gap width $L$ is important for $\xi$.  % Evaluating $\xi$
With the numbers given above, we find the axion Compton length
\begin{equation}
\lambdabar_a = {\frac {\hbar}{m_ac}} = 0.0197\,{\rm m},
\end{equation}
and keeping $L$ arbitrary, we get
\begin{equation}
\xi = % \frac{a_0m_aL}{\pi^2} = 2.64\times 10^{-4}L~ (\rm{eV})^2,
{\frac {\theta_0}{\pi^2}}\,{\frac {L}{\lambdabar_a}} = 1.3\times 10^{-18}\,{\frac {L}{{\rm m}}}.
\end{equation}
This estimate gives useful information: it shows that for astrophysical distances the relevant values for $L$ are easily compatible with  the condition of weak axion fields, $\xi \ll 1$. Our expression (\ref{maxrotation}) is thus applicable as it stands, for the mean Universe.

Let us now turn to the case of the neutron star polar cap region, studied by Noordhuis {\it et al.} \cite{noordhuis23,noordhuis24}, for which the static magnetic field is assumed to be very large, $B_0 = 10^{8}\,$G. For definiteness let us choose the axion energy density $\rho_a(\rm cap)$ region to be $10^{10}~$ times the mean-Universe value $\rho_a$ studied above. That gives for the axion field value the expression 
\begin{equation}
a_0{(\rm cap)} = 10^5 a_0,\qquad \theta_0(\rm cap)= 10^5 \theta_0,  \label{44}
\end{equation}
which in turn leads to
\begin{equation}
\xi{(\rm cap)} = % \frac{6.8\times 10^{10}}{L\rm [cm]}.
1.3\times 10^{-13}\,{\frac {L}{{\rm m}}}.
\end{equation}
The condition $\xi \ll  1$ can of course still be satisfied, but one might ask: does the opposite case $\xi \gg 1$ have a physical meaning?  That is apparently not so, however. Some insight is obtained from rewriting Eqs.~(\ref{exligning}) and (\ref{exligning1}) in the form
\begin{equation}
E_x'''' + [\lambda_n^2 E_x'' +i\mu \lambda_n^4\xi_n E_y] = \mu^2\lambda_n^4\xi_n^2 E_x,
\end{equation}
which shows that the rotational (i.e., imaginary)  term is proportional to $\xi_n$ while the dominant term on the right hand side is proportional to $\xi_n^2$. For large $\xi_n$ the rotation becomes very small.

\begin{figure}[t]
	\includegraphics[width=0.5\textwidth]{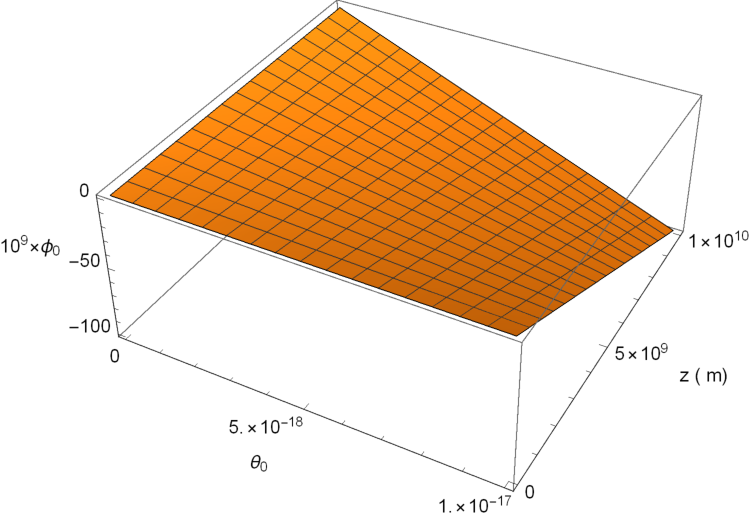}
	\caption{Variation of the reduced rotation angle $10^9\times \phi _0\left(\theta _0,z\right)$ as a function of $\theta _0\in \left[10^{21},10^{-17}\right]$ and of $z\in \left[10^5,10^{10}\right]$ m.}
	\label{fig1}
\end{figure}

As for the rotation angle $\phi(z)$ we obtain from Eq.~(\ref{maxrotation})
\begin{equation}\label{phiz}
\phi(z)= -\,{\frac{\mu}{2\pi}}\theta_0(\rm cap)z=-\,10^5\times \frac{\mu}{2\pi}\times \theta_0\times z,
\end{equation} 
where we have inserted the neutron star cap value for $\theta_0$. This expression is actually independent of the spacing $L$; a property that could have hardly been guessed in advance. By introducing the notation $\phi_0(z)=(2\pi/\mu)10^{-5}\phi(z)$ (the reduced rotation angle), we obtain $\phi_0(\theta_0,z)=-\,\theta_0z$. The variation of the reduced rotation angle  $\phi_0(\theta_0,z)$ as a function of $\theta _0$ and $z$,  is represented in Fig.~\ref{fig1}.

%%%%%%%%%%%%%%%%%%%%%%%%%%%%%%%%%%%%%%%%%%%%%%%%%%%%%%%%%%%%%%%%%%%%%%%%%%%%%%%%%%%%%%%%%%%%%%%%%%%%%%%%%
\subsection{Refilling of a vacuum gap in the polar cap region}
%%%%%%%%%%%%%%%%%%%%%%%%%%%%%%%%%%%%%%%%%%%%%%%%%%%%%%%%%%%%%%%%%%%%%%%%%%%%%%%%%%%%%%%%%%%%%%%%%%%%%%%%%

As our final point, we will consider the local `vacuum' gap regions in the neighborhood of the neutron star's polar axis, envisaged by Noordhuis {\it et al.} \cite{noordhuis23,noordhuis24}. A gap of this type, temporarily emptied of axions, is thought to be relatively small, typically a cylinder of radius 150 m and height 10 m. As is known, \cite{prabhu21}, the electromagnetic fields in the cap region are strong enough to produce  a large flux of axions. In turn, these axions may produce resonances, causing a broadband radio flux. Such local processes may make it natural to ask: what are the typical time scales involved? Let us make the simple estimate: if a vacuum gap exists for some time, what is the refilling time for the surrounding axion cloud to close the gap?

Let us start from the governing equation for the axions,
\begin{equation}\label{axionequation}
{\frac 1{c^2}}\partial_t^2 a - \nabla^2 a + \lambdabar_a^{-2} a
= -\,\sqrt{\frac {\varepsilon_0}{\mu_0}}\,{\frac {g_{a\gamma\gamma}}{\hbar c}}\,\bm{E}\cdot\bm{B}. 
\end{equation}
 For the magnetic field we insert $B_0= 10^{8}\,$T, and for the electric field we take the maximum value $E_0 = 6\times 10^{-6}cB_0$, as mentioned above. The magnetic field energy density is thus $W_{\rm magn} =  {\frac 1{2\mu_0}}B_0^2 = 3.98\times 10^{21}\,{\rm J/m}^3$. % (1/8\pi)10^{24}~$ erg/cm$^3$ = $2.49 \times 10^{25}~$ GeV/cm$^3$.
Multiplying with the electric field we obtain
\begin{equation}\label{50}
E_0B_0= 6\times 10^{-6}cB_0^2 = 12 \times 10^{-6} \sqrt{\frac {\mu_0}{\varepsilon_0}} W_{\rm magn}.
\end{equation}
In Eq.~(\ref{axionequation}) we maintain only the first term $\partial_t^2 a$ (thus omit the wave nature of the axion flow), and take the right hand side positive to adhere with the axion filling-state picture. This approximation is justified by the assumption of a thin axion filled cavity, which allows to neglect the impact of inhomogeneities on the wave field, and to apply the standard, non-dispersive wave equation to describe the axion field. Moreover, when considering the propagation of complex scalar waves, the terms proportional to the gradient of the axion field value can be ignored when the wavelength $\lambda \longrightarrow 0$. Hence, in this approximation, Eq.~(\ref{axionequation}) reduces to the eikonal equation, with the main focus  on the trajectory of the wave, rather than axion field value variations, which are assumed negligibly small in the thin cavity located in the polar cap of the neutron star.

 Thus we get the simplified equation
\begin{equation}
{\frac 1{c^2}}\partial_t^2 a = {\frac {g_{a\gamma\gamma}}{\hbar c}}\,\times (1.2\times 10^{-5})\,W_{\rm magn}
= 3.095\times 10^{14}\,{\frac 1{{\rm m}^3}}.
 \end{equation}
In the process of computations, we move between geometric and physical units by making use the conversion factor 1\,m\,$ = 5.068\times 10^{15}~{\rm GeV}^{-1}$. That leads to
\begin{equation}
        \partial_t^2 a = 5.45\times 10^{15}\,{\frac {\rm GeV}{s^2}}.
\end{equation}
Now integrate this equation over $t$, from $t=0$ to the filling time $\tau$, corresponding to the axion field value increasing from $a=0$ to the complete filling state $ a({\rm cap})$. This gives
\begin{equation}
a({\rm cap}) = {\frac {5.45}2}\times 10^{15}\,{\rm GeV}\,{\frac {\tau^2}{s^2}}
\end{equation}
Assuming $a_0({\rm cap}) = 10^5a_0 = 2.6\times 10^{-2}~$GeV as in Eq.~(\ref{44}) above, we obtain for the filling time
\begin{equation}
\tau = 3.09\times 10^{-9}\,{\rm s}.
\end{equation}
This is about what we would expect, although we ought to bear in mind the uncertainty in the estimate of $a_0({\rm cap})$.

%%%%%%%%%%%%%%%%%%%%%%%%%%%%%%%%%%%%%%%%%%%%%%%%%%%%%%%%%%%%%%%%%%%%%%%%%%%%%%%%%%%%%%%%%%%%%%%%%%%%%%%%%
\subsection{Observational implications}
%%%%%%%%%%%%%%%%%%%%%%%%%%%%%%%%%%%%%%%%%%%%%%%%%%%%%%%%%%%%%%%%%%%%%%%%%%%%%%%%%%%%%%%%%%%%%%%%%%%%%%%%%

The presence of specific modifications of the structure of the electromagnetic field in axion clouds photon provides the most direct possibility to show the existence of axions, namely, via  radio emission from the neutron star (pulsar) surface \cite{prabhu21,noordhuis23,noordhuis24}. For a neutron star evolving to an  equilibrium state, the emitted radio flux and spectrum will remain nearly constant until the  begins to spin down. Generally, the energy flux generated by the electromagnetic waves $(\bm{E},\bm{B})$  is given by the Poynting vector $\bm{S}={\frac 1{\mu _0}}\bm{E}\times \bm{B}$. The magnitude (absolute value) of the Poynting vector is given by $\left|\bm{S}\right| =S= {\frac 1{\mu_0}}\left|\bm{E}\right|\left|\bm{B}\right|\sin \alpha$, where $\alpha$ is the angle between the electric and magnetic field vectors $\bm{E}$ and $\bm{B}$. Since in an electromagnetic wave $\bm{E}\perp \bm{B}$, it follows that $\alpha =\pi/2$, and consequently generally $S={\frac 1{\mu_0}}\left|\bm{E}\right|\left|\bm{B}\right|= {\frac 1{\mu_0}}EB$. In the specific case of the electromagnetic energy emission from the axion dominated cavity, we can estimate the Poynting vector with the help of Eq. (\ref{50}) as given by $S= {\frac 1{\mu_0}}E_0B_0=1.2\times 10^{-7}\times \left(\varepsilon _0\mu _0\right)^{-1/2}W_{\rm magn}$, or $S\approx 1.42\times 10^{23} \;{\rm J/m^2\;s}$, where $c=1/\sqrt{\varepsilon_0 \mu_0}$. 

Assuming a circular emission area having a radius of $R_A=150$ m,  the electromagnetic luminosity of the axion dominated cavity is $L=4\pi R_A^2S\approx 4.05\times 10^{28}\; {\rm W}$.  The radio flux from the neutron star as observed on Earth can be estimated as $F=L/4\pi d^2$, where $d$ is the distance to the neutron star. For a source located at a distance of $d=100 $ kiloparsec, the received flux from the axion dominated cavity can be estimated as $F\approx 3.40\times 10^{-16}\; {\rm W/m^2}$. For the sake of comparison we would like to mention that the average Solar flux (irradiance) on the Earth surface is of the order of 1000 W/m$^2$. 

The sensitivity of a radio telescope can be estimated with the help of the parameter $S_e=F/\Delta \nu$, where $\Delta \nu$ is the signal bandwidth \cite{noordhuis24}. For a signal bandwidth of the order $\Delta \nu =250$ MHz, we obtain for the sensitivity of a radio telescope with respect to the energy flux from the axion gap the value $S_e=1.35\times 10^{-24} \; {\rm W/m^2 Hz}\approx 135$\,Jy. Hence, at least in principle, the signals coming from the axion filled gaps in neutron stars could be detected by recently designed radio telescopes, like, for example, LOFAR \cite{lofar}.     

In order to give a qualitative estimate of the frequency of the signal we approximate it as the axion frequency $\omega _a=m_ac^2/\hbar$. By assuming for the axion mass a range $0.7\times 10^{-5}\,{\rm eV}/c^2 < m_a < 3\times 10^{-5}\,{\rm eV}/c^2$, the frequency range of the radiation is given by $1.69\,{\rm GHz}< \nu _a < 7.24$ GHz. These values would reduce the required radio telescope sensitivity to the range $4.7\times 10^{-2}\,{\rm Jy} < S_e < 0.2$ Jy, with the corresponding signals still detectable by LOFAR.

%%%%%%%%%%%%%%%%%%%%%%%%%%%%%%%%%%%%%%%%%%%%%%%%%%%%%%%%%%%%%%%%%%%%%%%%%%%%%%%%%%%%%%%%%%%%%%%%%%%%%%%%%
\section{Summary}
%%%%%%%%%%%%%%%%%%%%%%%%%%%%%%%%%%%%%%%%%%%%%%%%%%%%%%%%%%%%%%%%%%%%%%%%%%%%%%%%%%%%%%%%%%%%%%%%%%%%%%%%%

The possible existence of a dense and locally inhomogeneous axion region in the polar cap of a neutron star is of obvious physical interest \cite{noordhuis23,noordhuis24}, and has been the main motivation for our present investigation. In this picture of a neutron star, there is a strong static magnetic field $B_0 \sim 10^{8}\,$T as well as a strong electric field $E_0 \sim 10^{-6}cB_0$, both directed  orthogonally to star's surface. 

In our theoretical description we assumed conventional values for the axion mass, $m_a = 10^{-5}\,{\rm eV}/c^2$, as well as for the maximum  $a_0 = 260~$eV of  the harmonically varying axion $a$. Thus, in this description the dense property of the axion cloud is due to the high {\it number} of axions, and not attributed to  individual high energy. A further strong motivation for our analysis is that the idea of using metallic haloscopes to detect axions (cf., for instance, Ref.~\cite{sikivie83} and further references listed above), has turned out to b extremely challenging. In the so-called `antenna' experiments a magnetic field of $B_0=10~$T, a strong field under terrestrial conditions, will according to theory only be able to create an electromagnetic flux of about one photon per day \cite{liu22}. 

In our analysis, after briefly surveying these points, we presented the basic equations for the mixed electromagnetic-axion field, emphasizing the hybrid formulation (Sec.~\ref{fieldeqs2}) in which the axion terms appear explicitly as source terms. Our basic geometrical model, of parallel-plates Casimir type, was set up in Sec.~\ref{model}, containing a uniform spatial inhomogeneity in the axion density distribution, and we calculated to the lowest order  the rotation angle of the polarization plane. In principle, this is a measurable quantity. It is noteworthy that this method only works when the axions influence in the formalism is not too strong. 

One should also mention that not only spatial inhomogeneity, but also a time dependence of the axion field, may result in polarization plane rotation, although we did not consider the latter aspect in this theory. Finally, we made an estimate of  the filling time $\tau$ of the temporary `voids' reported by Noordhuis {\it et al.}, and found $\tau$  to be of order nanoseconds, which is within the accuracy of atomic clocks precision and thus detectable.

%%%%%%%%%%%%%%%%%%%%%%%%%%%%%%%%%%%%%%%%%%%%%%%%%%%%%%%%%%%%%%%%%%%%%%%%%%%%%%%%%%%%%%%%%%%%%%%%%%%%%%%%%
\begin{acknowledgments}
We would like to thank the anonymous reviewer for comments and suggestions that helped us to improve our work. We are grateful to Markku Oksanen and Anca Tureanu for illuminating discussions. The work of YNO was partialy supported by the Russian Science Foundation Grant 26-79-32004.
\end{acknowledgments}
%%%%%%%%%%%%%%%%%%%%%%%%%%%%%%%%%%%%%%%%%%%%%%%%%%%%%%%%%%%%%%%%%%%%%%%%%%%%%%%%%%%%%%%%%%%%%%%%%%%%%%%%%

% Set the ending of a LaTeX document
\end{document}